\documentclass[12pt]{article}
\usepackage{amsmath}
\usepackage{amssymb}%mathbb
\usepackage{graphicx}
\usepackage{indentfirst}% 僇邈忑俴坫輛

\newtheorem{theorem}{Theorem}

\makeatletter
\def\ExtendSymbol#1#2#3#4#5{\ext@arrow 0099{\arrowfill@#1#2#3}{#4}{#5}}
\def\RightExtendSymbol#1#2#3#4#5{\ext@arrow 0359{\arrowfill@#1#2#3}{#4}{#5}}
\def\LeftExtendSymbol#1#2#3#4#5{\ext@arrow 6095{\arrowfill@#1#2#3}{#4}{#5}}
\makeatother

\begin{document}
\baselineskip 20pt

\title{Ascertaining the Uncertainty Relations via Quantum Correlations}

\author{Jun-Li Li, Kun Du, and
Cong-Feng Qiao\footnote{Corresponding author: qiaocf@ucas.ac.cn}\\[0.5cm]
\small School of Physics, University of Chinese Academy of Sciences \\
\small YuQuan Road 19A, Beijing 100049, China\\[0.2cm]
}

\date{}
\maketitle

\begin{abstract}

We propose a new scheme to express the uncertainty principle in form of
inequality of the bipartite correlation functions for a given multipartite
state, which provides an experimentally feasible and model-independent way
to verify various uncertainty and measurement disturbance relations. By
virtue of this scheme the implementation of experimental measurement on
the measurement disturbance relation to a variety of physical systems
becomes practical. The inequality in turn also imposes a constraint on the
strength of correlation, i.e. it determines the maximum value of the
correlation function for two-body system and a monogamy relation of the
bipartite correlation functions for multipartite system.
\end{abstract}

%\hspace{3mm} {\bf Key words:} Uncertainty relation, Correlation
%function, Quantum nonlocality.

The uncertainty principle lies at the heart of quantum mechanics and is one
of the most fundamental features which distinguish it from the classical
mechanics. The original form, $p_1q_1\sim h$, stems from a heuristic
discussion of Heisenberg on Compton scattering \cite{Heisenberg-o} where
$p_1$, $q_1$ are the determinable precisions of position and momentum, $h$ is
the Planck constant. A generalization to arbitrary pairs of observables is $
\Delta {A} \Delta {B} \geq |\langle [A,B] \rangle|/2 $, where the standard
deviation is $\Delta {X} = (\langle X^2\rangle - \langle X \rangle^2)^{1/2}$,
$X =A\ \text{or}\ B$, $\langle \cdots \rangle$ stands for expectation value,
and the commutator is defined as $[A,B] \equiv AB - BA$. This is the usually
called Heisenberg-Robertson uncertainty relation \cite{Robertson}. A more
stronger version is the Robertson-Schr\"odinger uncertainty relation
\cite{Schrodinger} which takes the form of $(\Delta {A})^2 (\Delta {B})^2
\geq (\langle \{A, B\} \rangle/2 - \langle A \rangle \langle B \rangle )^2 +
|\langle [A, B] \rangle|^2/4 $ where the anticommutator is defined as
$\{A,B\} \equiv AB + BA$.

Note that in the form involving standard deviations, the uncertainty
relation represents the property of the ensemble of arbitrary quantum state
in Hilbert space and does not concern with the specific measurements. Thus
such uncertainty relation is not related to the precision of measurement on
one observable and the disturbance to its conjugate.

If we assume $\epsilon(A)$ to be the precision of the measurement on $A$ and
$\eta(B)$ to be the disturbance of the same measurement on $B$, the
Heisenberg-type relation with regard to measurement and disturbance would
read
\begin{eqnarray}
\epsilon(A) \eta(B) \geq \frac{1}{2}| \langle [A, B] \rangle | \; . \label{H-MDR}
\end{eqnarray}
In recently, Ozawa found that the this form of measurement disturbance
relation (MDR) (\ref{H-MDR}) is not a universal one, and a new MDR was
proposed \cite{Ozawa-operator}, which are thought to be generally valid,
i.e.
\begin{eqnarray}
\epsilon(A)\eta(B) + \epsilon(A)\Delta B + \Delta A \, \eta(B)
\geq \frac{1}{2} |\langle [A, B] \rangle|
\; . \label{O-MDR}
\end{eqnarray}
Eq.(\ref{O-MDR}) is of fundamental importance, for example, it leads to a
totally different accuracy limit $\epsilon(A)$ for non-disturbing
measurements ($\eta(B)=0$) comparing to the Heisenberg-type MDR. In quantum
information science, the uncertainty principle in general is also crucial to
the security of certain protocols in quantum cryptography \cite{QKD-UP}, and
additionally, it plays an important role in the quantum metrology
\cite{quantum-metrology}.

Despite the importance of the uncertainty principle, only the uncertainty
relation in form of standard deviations has been well verified in various
situations, e.g., see \cite{Standard-uncertainty-exp} and the references
therein. Experiments concerning both Heisenberg-type and Ozawa's MDRs have
just been performed with neutrons \cite{MDR-Neutron} and photons
\cite{MDR-Photon}. For neutrons in a given polarization state, the error and
disturbance can be statistically determined based on a method proposed by
Ozawa \cite{Method-Ozawa}. In the photon experiment, the weak measurement
model introduced in \cite{MDR-Weak-values} was employed for the measurement.
Large samples of data is necessary due to the sensitivity to the measurement
strength of a weak measurement process which is used for gathering
information of the system prior to the actual measurement
\cite{Polarization-Weak-values}. The results of \cite{MDR-Neutron} and
\cite{MDR-Photon} exhibit the validation of Ozawa's MDR but rather the
Heisenberg-type. Since the uncertainty principle limits our ultimate ability
to reduce noise when gaining information from the state of a physical
system, its experimental verification in various systems and different
measurement interactions is still an important subject.

Here in this work, we present such a general scheme from which both the
uncertainty relation and MDR turn to the forms involving only bipartite
correlation functions. In this formalism, whilst the uncertainty relation
becomes an inequality imposed on the correlation functions of bipartite
states, the different forms of MDRs transform into strong constraints on the
shareability (monogamy) of the bipartite correlations in multipartite state.
This directly relates the key element of quantum information, i.e., the
nonlocal correlation, with the fundamental principle of quantum mechanics,
i.e., uncertainty principle, in a quantitative way. And most importantly, it
enables us to test the MDRs in a variety of physical systems.

To test the validity of the various MDRs, one has to measure the physical
observable quantities for which the different MDRs exhibit distinct
responses. Here we present our method of constructing such quantities for
qubit systems. Although the generalization to arbitrary systems is not
trivial, the various MDRs have already shown the essential differences in
two-dimensional Hilbert spaces within our scheme. The qubit systems include
spin 1/2 particle, polaizations of photons, two level atoms, etc. For the
sake of convenience we take the measurable observables to be the spin
components. A measurement of spin along arbitrary vector $\vec{a}$ in three
dimensional Euclidean space can be represented by the following operator
\begin{eqnarray}
A = \vec{\sigma} \cdot \vec{a} = |\vec{a}| \vec{\sigma} \cdot \vec{n}_a
\; . \label{operator-def}
\end{eqnarray}
Here $\vec{\sigma} = (\sigma_x, \sigma_y, \sigma_z)$ are Pauli matrices,
$\vec{n}_a = \vec{a}/|\vec{a}|$, and a general commutative relation holds for
such operators
\begin{eqnarray}
[A, B] = 2iC \; ,  \label{basic-commutator}
\end{eqnarray}
where $B = \vec{\sigma}\cdot \vec{b}$, $C = \vec{\sigma} \cdot \vec{c}$,
$\vec{c} = \vec{a}\times \vec{b}$. Let $|n^{\pm}_p\rangle$ be the two
eigenvectors of operator $P = \vec{\sigma} \cdot \vec{n}_p$ with eigenvalues
$\pm1$, the following complete relations hold
\begin{eqnarray}
|n^+_p\rangle \langle n^+_p| + |n^-_p\rangle \langle n^-_p|= 1 \; , \;
|n^+_p \rangle \langle n^+_p| - |n^-_p \rangle \langle n^-_p| =
\vec{\sigma} \cdot \vec{n}_p = P \; . \label{complete-relation}
\end{eqnarray}
Here $\vec{n}_p$ is a unit vector, $|n_p^{\pm}\rangle \langle n_{p}^{\pm}|
\equiv P^{\pm}$ are the projection operators. Using the Schmidt
decomposition, any bipartite pure state is unitarily equivalent to the state
\cite{QIP-Book}: $|\psi_{12} \rangle = \alpha |+\rangle|+\rangle + \beta|-
\rangle|-\rangle$ where $|\alpha|^2 + |\beta|^2 =1$, and $\alpha\geq 0$,
$\beta\geq 0$. The correlation function between two operators $A$ and $B$
for arbitrary quantum state $|\psi\rangle$ is defined as $E(A_1,B_2) =
\langle \psi|A_{1}\otimes B_2|\psi\rangle$. Here the subscripts of $A$, $B$
stand for the corresponding partite which they are acting.

For the Robertson-Schr\"odinger uncertainty relation we have the following
theorem:
\begin{theorem}
The Robertson-Schr\"odinger uncertainty relation imply the following
inequality on the correlation functions of arbitrary bipartite quantum state
\begin{eqnarray}
\left| E(A_1, P_2)\vec{b} - E(B_1, P_2)\vec{a} \right|^2 +
\left|E(C_1, P_2) \right|^2 \leq S^2 \; , \nonumber
\end{eqnarray}
where $X_i = \vec{\sigma}_i \cdot\vec{x}$, $X = A,\ B,\ \text{or}\ C$,
$\vec{c} = \vec{a} \times \vec{b}$, $P_i=\vec{\sigma}_i \cdot \vec{n}_p$,
$\vec{n}_p$ is unit vector, $i=1,2$ denote the corresponding partite, $S$ is
the parallelogram area formed by $\vec{a}$, $\vec{b}$. \label{theorem-R-S}
\end{theorem}

This theorem indicates that the correlation functions between one specific
operator ($P$) and two other operators ($A$, $B$) and their commutator ($C$)
in bipartite states are constrained by the area of parallelogram formed with
$\vec{a}$ and $\vec{b}$. The maximal attainable value of the bipartite
correlation function is $E(A_1, A_2) = |\vec{a}|^2$ which is the area of a
square with length $|\vec{a}|$. A proof of this theorem is given in Appendix
A.

\begin{figure} \centering
\scalebox{0.5}{\includegraphics{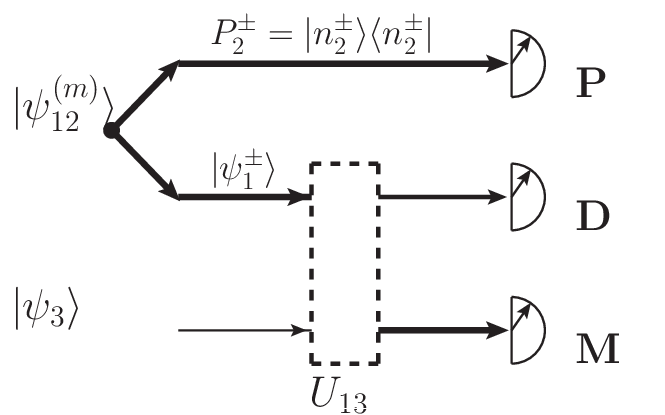}}
\caption{Illustration of the detection of measurement precision and disturbance.
{\bf P}, {\bf D}, {\bf M} stand for the function of projection, disturbance,
and measuring. A meeter system $|\psi_{3}\rangle$ interacts with the signal state
$|\psi_{1}^{\pm}\rangle$ which is prepared by projecting a bipartite entangled state
$|\psi_{12}^{(m)}\rangle$ at {\bf P}. The measurement result
can be obtained from {\bf M}, and the measurement disturbance on signal
$|\psi_1^{\pm}\rangle$ will be detected at {\bf D}. }
\label{Fig-Measure-PDM}
\end{figure}

As for the MDR, it is a subtle problem in quantum theory. In order to detect
the influence (disturbance) on quantity $B$ introduced in measuring $A$, one
needs to measure $B$ before and after the measurement on $A$. If the initial
state is not $B$'s eigenstate, the acquisition of information on $B$ prior
to the measurement $A$ will inevitably change the the initial state and
makes the subsequent measurement process irrelevant to the initial state. To
illustrate this, a simple measurement scheme is presented in
Fig.\ref{Fig-Measure-PDM} where the measurement is performed via the
interaction of the signal system $|\psi_{1}^{\pm}\rangle$ with a meter
system $|\psi_3\rangle$ \cite{MDR-Weak-values}.

The Ozawa's precision and disturbance quantities in Eq.(\ref{O-MDR}) are
defined as \cite{Ozawa-operator}
\begin{eqnarray}
\epsilon(A)^2 & \equiv & \langle [U^{\dag}_{13}(I_1 \otimes M_3) U_{13} -
A_1\otimes I_3 ]^2 \rangle \; , \label{def-precision} \\
\eta(B)^2 & \equiv &  \langle [U^{\dag}_{13}(B_1 \otimes I_3) U_{13} -
B_1\otimes I_3]^2 \rangle \; . \label{def-disturbance}
\end{eqnarray}
Here the expectation values in Eqs.(\ref{def-precision},
\ref{def-disturbance}) are evaluated with the same compound state
$|\psi_{1}\rangle |\psi_{3}\rangle$, where $|\psi_1\rangle$ can be
arbitrary, i.e., $|\psi_1^{\pm}\rangle$; $|\psi_3\rangle$ is the quantum
state of the measurement apparatus; $U_{13}$ is a unitary measurement
interaction. If the measurement process is carried out via spin dependent
interaction with a qubit state (partite 3) and regarding the measurement
read out of the spin of partite 3 to be the measurement result of the signal
state $|\psi_1\rangle$, we can have $M_3 \to A_3$. It is obvious that in
determining $\eta(B)$ (Eq.(\ref{def-disturbance})), we have to measure $B_1$
before and after the measurement interaction $U_{13}$.

Our procedure to settle the measurement problem under Ozawa's defintions
goes as follows. Suppose we want to measure the MDR with respect to any
given pair of spin components of $A_1 = \vec{\sigma}_1\cdot \vec{a}$ and
$B_1 = \vec{\sigma}_1 \cdot \vec{b}$ for arbitrary state $|\psi_1\rangle$.
This state can be prepared via the following entangled state
\begin{eqnarray}
|\psi_{12}^{(m)}\rangle = \frac{1}{\sqrt{2}}
\left( |+\rangle_c|-\rangle_{c} + (-1)^m |-\rangle_c|+\rangle_c \right) \; .
\end{eqnarray}
Here, $m\in \{0,1\}$; $\vec{c} = \vec{a} \times \vec{b}$ and
$|\pm\rangle_{c}$ are the spin eigenstates along $\vec{c}$ ($|\pm\rangle$
stand for the eigenstates along $z$ if not specified). Without loss of
generality, we can set the $\vec{a}$-$\vec{b}$ plane as $x$-$z$ plane then
$\vec{c}$ is along the $y$ axis
\begin{eqnarray}
|\psi_{12}^{(1)}\rangle & = & \frac{1}{\sqrt{2}}
\left( |+-\rangle - |-+\rangle \right) \; , \\
 |\psi_{12}^{(0)}\rangle &=& \frac{1}{\sqrt{2}}
 \left( |++\rangle + |--\rangle \right) \; .
\end{eqnarray}
$|\psi_{12}^{(m)}\rangle$ have the following property
\begin{eqnarray}
V_1\otimes V^{-1}_2|\psi_{12}^{(m)}\rangle =
(-1)^{m}|\psi_{12}^{(m)}\rangle \; , \; m \in\{0,1\} \; ,
\label{rotation-invariant-12}
\end{eqnarray}
where $V_i =\vec{\sigma}_i \cdot\vec{v}$ is an operator acting on the $i$th
partite and $\vec{v}$ is a unit vector in the $\vec{a}$-$\vec{b}$ (i.e.,
$x$-$z$) plane. With the definition of projection operators in
Eq.(\ref{complete-relation}), an arbitrary quantum state ($|\psi_1\rangle$)
of partite 1 can be obtained via a projective measurement {\bf P} on partite
2 (see Fig.\ref{Fig-Measure-PDM})
\begin{eqnarray}
|\psi_{1}^{\pm}\rangle & = &
\frac{_2\langle n_p^{\pm} |\psi_{12}^{(m)} \rangle}{|_2\langle n_p^{\pm}
|\psi_{12}^{(m)}\rangle|} \; .
\label{project-to-psi1}
\end{eqnarray}
Here in the present situation $|_2\langle n_p^{\pm}|\psi_{12}^{(m)}\rangle|
= 1/\sqrt{2}$ and the arbitrariness of $|\psi_{1}^{\pm}\rangle$ is
guaranteed by the arbitrariness of $\vec{n}_{p}$.

The measurement precision of quantity $A$ for quantum state
$|\psi_{1}^{\pm}\rangle$ and the corresponding disturbance on another
quantity $B$ now can be written as
\begin{eqnarray}
\epsilon^{\pm}(A)^2 = \langle \psi_3|\langle \psi_1^{\pm}
|\left[ U_{13}^{\dag}(I_1\otimes A_3)U_{13} -
A_1 \otimes I_3 \right]^2 |\psi_1^{\pm}\rangle |\psi_3\rangle \; , \\
\eta^{\pm}(B)^2 = \langle \psi_3|\langle \psi_1^{\pm}|\left[ U_{13}^{\dag} ( B_1
 \otimes I_3 )U_{13} -
B_1 \otimes I_3 \right]^2 |\psi_1^{\pm}\rangle |\psi_3\rangle \; .
\end{eqnarray}
With these definitions, we can derive the following relation (see the
Appendix B)
\begin{eqnarray}
& & |\vec{a}|^2 + |\vec{b}|^2 - (-1)^m [E(A_2,A_3) + E(B_1, B_2)] \nonumber \\
& = & \frac{1}{4} \left[ \epsilon^+(A)^2 + \eta^+(B)^2 +
\epsilon^-(A)^2 + \eta^-(B)^2 \right] \; , \label{MDR-Correlation}
\end{eqnarray}
where the correlation function $E(X_i, X_j) = \langle \psi_{123}|X_i \otimes
X_j|\psi_{123}\rangle$, $X = A\ \text{or}\ B$, $|\psi_{123}\rangle \equiv
U_{13}|\psi_{12}^{(m)}\rangle |\psi_{3}\rangle$, $i,j \in \{1,2,3\}$, the
subscripts of operators stand for the corresponding partite which they are
acting. The precision and disturbance of the measurement now are directly
related to the bipartite correlation functions of a tripartite state.
Eq.(\ref{MDR-Correlation}) is universally valid regardless of the
measurement interaction $U_{13}$ which brings about the tripartite state.

For arbitrary given state $|\psi_1^{\pm}\rangle$, the Heisenberg-type and
Ozawa's MDRs read
\begin{eqnarray}
\epsilon^{\pm}(A)\eta^{\pm}(B) \geq \frac{1}{2}
|\langle \psi_1^{\pm}| [A,B] |\psi_{1}^{\pm}\rangle| \; , \label{Heisenberg-MDR-PD} \\
 \epsilon^{\pm}(A)\eta^{\pm}(B) + \epsilon^{\pm}(A)\Delta^{\pm}(B) + \eta^{\pm}(B)
\Delta^{\pm}(A) \geq \frac{1}{2}|\langle \psi_{1}^{\pm} |[A, B]|\psi_{1}^{\pm} \rangle|
\; . \label{Ozawa-MDR-PD}
\end{eqnarray}
An intuitive view of the above equations tells that the allowed regions for
$\epsilon$ and $\eta$ lie above the hyperbolic curves of $\epsilon^{\pm}(A)$
and $\eta^{\pm}(B)$ in the quadrant I. The constraints
Eqs.(\ref{Heisenberg-MDR-PD},\ref{Ozawa-MDR-PD}) are then transferred to the
bipartite correlation functions via Eq.(\ref{MDR-Correlation}). Thus we have
the following theorem
\begin{theorem}
For $A=\vec{\sigma} \cdot \vec{a}$, $B = \vec{\sigma} \cdot \vec{b}$, a
tripartite state can be obtained by interacting one partite of
$|\psi_{12}^{(m)}\rangle$ with a third partite 3. The Heisenberg-type and
Ozawa's MDRs imply the following different relations on the resulted
tripartite state
\begin{eqnarray}
E(A_2, A_3) + E(B_1, B_2)
\leq |\vec{a}|^2 + |\vec{b}|^2 - \kappa_{h,o}|\vec{n}_p \cdot (\vec{a}\times\vec{b})| \; .
\label{upper-Heisenberg-Ozawa}
\end{eqnarray}
Here $E(X_i, X_j)$ are the bipartite correlation functions of the tripartite
state, $\kappa_{h} = 1$ and $ \kappa_o= (\sqrt{2}-1)^2$ for Heisenberg-type
and Ozawa's MDR respectively, $\vec{n}_p$ is an arbitrary unit vector.
\label{Theorem-Heisenberg-Ozawa}
\end{theorem}

The proof of Theorem \ref{Theorem-Heisenberg-Ozawa} is presented in
Appendix C. From Theorem \ref{theorem-R-S} we know that $|\vec{a}|^2$
and $|\vec{b}|^2$ are the maximum values of $E(A_2, A_3)$ and $E(B_1,
B_2)$ in bipartite states. Now due to Theorem
\ref{Theorem-Heisenberg-Ozawa} the maximum of the sum of the two
bipartite correlations in the tripartite state is reduced by an
amount proportional to the volume of the parallelepiped with edges
$\vec{a}$, $\vec{b}$, and $\vec{n}_p$.

\begin{figure}\centering
\scalebox{0.5}{\includegraphics{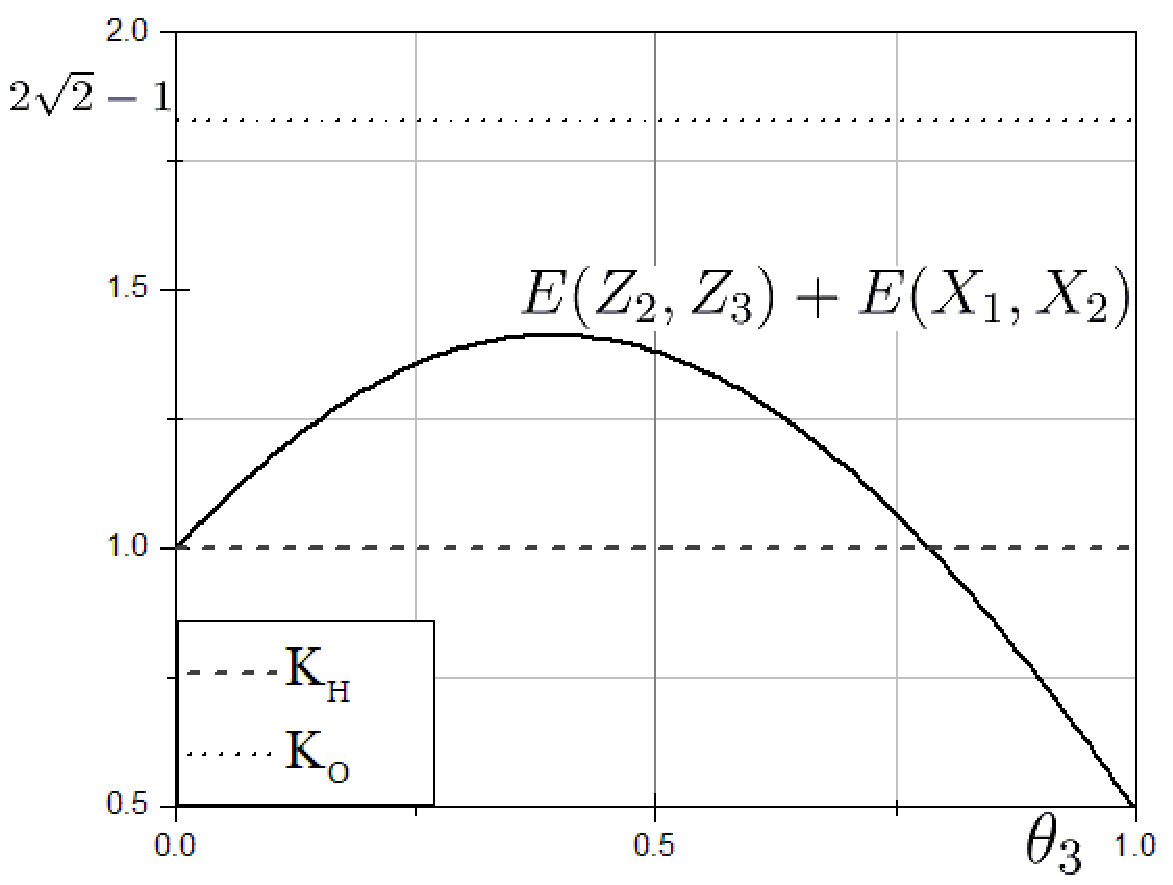}}\caption{The demonstration
of Heisenberg-type and Ozawa's MDR with measurement precision of $A = Z$
and its disturbance on $B = X$. Here $K_{H,O}$ are the upper bound imposed
by Heisenberg-type and Ozawa's MDR at $\theta_p=0$ respectively. The sum
$E(Z_2, Z_3) + E(X_1,X_2)$ surpasses the limit imposed by
Heisenberg-type MDR. }
\label{Fig-violation}
\end{figure}

The experiments to test the validity of the MDRs become straightforward due
to Theorem \ref{Theorem-Heisenberg-Ozawa}. Here we present an example of the
measurement model of qubit system with the measurement interaction $U_{13}$
being the CNOT gate \cite{MDR-Weak-values} within our method. Suppose we
want to measure the precision of $Z = \sigma_z$ and the disturbance on $X =
\sigma_x$ for an arbitrary qubit state $|\psi_{1}\rangle$. Following Theorem
\ref{Theorem-Heisenberg-Ozawa}, on choosing $|\psi_{12}^{(1)}\rangle =
\frac{1}{\sqrt{2}}(|++\rangle + |--\rangle)$, the measurement interaction
CNOT gate between one partite of $|\psi_{12}^{(1)}\rangle$ and the meeter
system $|\psi_3\rangle = \cos\theta_3|+\rangle + \sin\theta_3|-\rangle$ will
lead to the following tripartite state
\begin{eqnarray}
|\psi_{123}\rangle & = &
\frac{1}{\sqrt{2}}[|++\rangle(\cos\theta_3|+\rangle + \sin\theta_3|-\rangle) + \nonumber \\
& & \hspace{0.8cm} |--\rangle (\cos\theta_3|-\rangle + \sin\theta_3|+\rangle)] \; .
\end{eqnarray}
According to Theorem \ref{Theorem-Heisenberg-Ozawa}, the Heisenberg-type and
Ozawa's MDRs impose the following constraints on the bipartite correlation
functions of $|\psi_{123}\rangle$
\begin{eqnarray}
\text{Heisenberg-type MDR: }\; E(Z_2,Z_3) + E(X_1,X_2)
& \leq & 2-|\cos\theta_p| \; , \\
\text{Ozawa's MDR: }\;  E(Z_2, Z_3) + E( X_1, X_2)
& \leq & 2-(\sqrt{2}-1)^2|\cos\theta_p|  \; ,
\end{eqnarray}
for arbitrary $\theta_p$, the angle between $\vec{n}_p$ and $\vec{c}$. The
tightest bound happens when $\theta_p=0$. Thus a measurement of bipartite
correlation function of $E(Z_2, Z_3)$, $E(X_1, X_2)$ in the tripartite state
would be capable to verify the Heisenberg-type and Ozawa's MDR (see
Fig.\ref{Fig-violation}). That is the Heisenberg-type MDR will be violated
provided that the experimental result agrees with the solid line of $E(Z_2,
Z_3) + E(X_1, X_2)$ in Fig.\ref{Fig-violation}.

From the above example, the procedure of our scheme can be summarized as:
(1) prepare a bipartite entangled state, (2) interact one partite of the
entangled state with a third partite, and (3) measure the bipartite
correlation functions of the resulted tripartite state. The generation of
the bipartite entangled state has already been realized in various systems,
e.g. photons \cite{photon-1,photon-2}, atoms
\cite{atom-entanglement,atom-RMP}, and high energy particles
\cite{high-energy,high-energy-2}. The further interaction of one partite of
the entangled state with a third partite can also be arbitrary, i.e.,
elastic or inelastic collisions, or via optical cavities, etc. More
importantly, we need only to measure the bipartite correlation functions of
the obtained tripartite state rather than the measurement precision and
disturbance, which may not be easy to quantify for some types of measurement
interactions. Hence, our scheme could be applied to a large number of
systems in the verification of the MDRs.

In addition to a clear discrimination between the Heisenberg-type and
Ozawa's MDRs, a more important physical consequence of the Theorem
\ref{Theorem-Heisenberg-Ozawa} is that it reveals a monogamy relation on
Bell correlations \cite{Bell-inequality, CHSH-monogamy, Bell-monogamy} in
the tripartite entangled state. According to the Theorem
\ref{Theorem-Heisenberg-Ozawa}, when measuring the precision of $B$ and the
disturbance it imposes on $A$, we will have
\begin{eqnarray}
|E(B_2, B_3) + E( A_1, A_2)|
\leq |\vec{a}|^2 + |\vec{b}|^2 - \kappa_{h,o}|\vec{n}_p \cdot (\vec{a}\times\vec{b})|
 \; , \label{upper-Heisenberg2}
\end{eqnarray}
Introducing two new vectors $\vec{a}\,' = \frac{1}{2}(\vec{a} + \vec{b})$ ,
$\vec{b}' = \frac{1}{2}(\vec{b} - \vec{a})$, we can similarly define $A' =
\vec{\sigma} \cdot \vec{a}'$, $B' = \vec{\sigma} \cdot \vec{b}'$. Following
the definition of correlation function in Eq.(\ref{MDR-Correlation}), we can
get
\begin{eqnarray}
E(A_i, A_j) & = & E(A_i, A'_j) - E(A_i,B'_j) \; ,\label{Bell-tran1} \\
E(B_i, B_j) & = & E(B_i, A'_j) + E(B_i, B'_j) \; . \label{Bell-tran2}
\end{eqnarray}
Adding Eq.(\ref{upper-Heisenberg-Ozawa}) and Eq.(\ref{upper-Heisenberg2}),
and taking Eqs.(\ref{Bell-tran1},\ref{Bell-tran2}), we have
\begin{eqnarray}
& & \left| E(A_2,A'_3) - E(A_2,B'_3) + E(B_2, A'_{3})+ E(B_2, B'_{3})  + \right. \nonumber \\
& & \left. \hspace{0.1cm} E(A_1, A'_2)   - E(A_1, B'_2) +
E(B_1, A'_{2})  + E(B_1, B'_{2}) \hspace{0.2cm}\right|
\leq 2K_{H,O} \; . \label{Bell-monogamy}
\end{eqnarray}
where $K_{H,O}=|\vec{a}|^2 + |\vec{b}|^2 - \kappa_{h,o}|\vec{n}_p \cdot
(\vec{a}\times\vec{b})|$. When $|\vec{a}|=|\vec{b}|=1$, $\vec{a}
\perp\vec{b}$, Eq.(\ref{Bell-monogamy}) leads to the sum of two particular
CHSH type correlations \cite{CHSH}
\begin{eqnarray}
\left|B_{\text{CHSH}}^{(23)} + B_{\text{CHSH}}^{(12)}\right| \leq 2\sqrt{2} K_{H,O} \;  .
\end{eqnarray}
Here $B_{\text{CHSH}}^{(ij)} = E(A_i,A'_j) - E(A_i,B'_j) + E(B_i,A'_j) +
E(B_i,B'_j)$. The tightest bound also happens when $\theta_p=0$, which lead
the following
\begin{eqnarray}
\text{Heisenberg-type MDR: }\;
\left|B_{\text{CHSH}}^{(23)} + B_{\text{CHSH}}^{(12)}\right| & \leq & 2\sqrt{2} \; , \\
\text{Ozawa's MDR: }\;
\left|B_{\text{CHSH}}^{(23)} + B_{\text{CHSH}}^{(12)}\right| & \leq & 2\sqrt{2}(2\sqrt{2}-1)
\; .
\end{eqnarray}
The above monogamy relations on quantum nonlocality are direct results of
the MDRs according to our theorem. Note, there are also discussions in the
literature on Bell correlations based on the entropic measures of
uncertainty relation \cite{UP-determine-nonlocal, Complementary-monogamy}.

It should be noted that the definitions of measurement precision and
disturbance in Eqs.(\ref{def-precision},\ref{def-disturbance}) by Ozawa
involve the comparisons of the same physical observable before and after the
measurement, thus base on practical physical motivations. However, the exact
definitions that capture the full physical contents of the measurement error
and disturbance are still under study \cite{Busch-1, Dressel,Busch-2}.
Nevertheless, Ozawa's definitions and the resulted MDRs may be regarded as
one of the best attempts to capture the quantitative descriptions of the
measurement and its back action in quantum mechanics. The method we
presented just provides a powerful tool to study the physical consequences
of the MDRs which is meaningful in judging their usefulness. For example,
our method transforms the MDRs into inequalities of correlation functions of
tripartite entangled state. In this way the importance of the MDRs manifests
in their connections with the quantum entanglement which is a key physical
resource in quantum information science and has a close relation with
quantum metrology \cite{quantum-metrology}. Meanwhile, in principle the idea
of our scheme may also be applied to other definitions of the error and
disturbance. This would enable the method to examine the meaningfulness of
the variant definitions.

In conclusion, we proposed in this work a general scheme to express the
uncertainty principle in terms of bipartite correlation functions, by which
the essential differences between the MDRs are characterized by the
inequalities constraining the correlation functions of multipartite state.
This not only builds a bridge between the MDRs and the quantum entanglement
but also provides a way to study the direct physical consequences of such
fundamental relations. The resulted inequalities reveal that both the
strength and the shareability (monogamy) of the quantum correlation are
determined by the uncertainty principle. Further studies on the uncertainty
relation and MDRs with, e.g., atoms, ions, or even high energy particles
become possible due to our scheme. The connections between MDRs and
entanglement revealed in our scheme may also shed new light on the the
studies of the relations between the MDRs and the quantum cryptography,
quantum metrology, etc.

Note: after the completion of the manuscript, there has been some progress
in the study of MDRs, i.e., \cite{Weston-Pryde},\cite{Branciard},etc. Our
method may apply to such cases as well and these MDRs would also give
distinct constraints on quantum correlations \cite{Inprogress}.)

\vspace{1cm}
%%%%%%%%%%%%%%%%%%%%%%%%%%%%%%%%%%%%%%%%%%%%%%%%%%%%%%%%%%%%%%%%%%%%%%
\noindent {\bf Acknowledgments}

This work was supported in part by the National Natural Science Foundation
of China(NSFC) under the grants 10935012, 11121092, 11175249 and 11205239.
%%%%%%%%%%%%%%%%%%%%%%%%%%%%%%%%%%%%%%%%%%%%%%%%%%%%%%%%%%%%%%%%%%%%%%

\vspace{3cm}

%\newpage

\appendix{\noindent {\bf\Large Appendix}}

\section{Proof of theorem \ref{theorem-R-S}} \label{appendix-1}

Proof of the equation of theorem \ref{theorem-R-S}:

\begin{eqnarray}
\left| E(A_1, P_2)\vec{b} - E(B_1, P_2)\vec{a} \right|^2 +
\left|E(C_1, P_2) \right|^2 \leq S^2 \; . \nonumber
\end{eqnarray}

\noindent {\bf Proof}: Following the definition of the standard
deviation, the Robertson-Schr\"odinger uncertainty relation takes the
following form
\begin{eqnarray}
(\langle A^2\rangle  - \langle A\rangle^2) (\langle
B^2\rangle  - \langle B\rangle^2) \geq \left( \frac{1}{2}\langle
AB + BA \rangle -
\langle A\rangle \langle B \rangle \right)^2 +
\frac{1}{4} \left| \langle [A, B] \rangle \right|^2 \; .
\label{R-S-Relation-1}
\end{eqnarray}
With the definition of operators as in  Eq.(\ref{operator-def}) and the
basic commutator Eq.(\ref{basic-commutator}), Eq.(\ref{R-S-Relation-1}) can
be written as
\begin{eqnarray}
|\vec{a}|^2|\vec{b}|^2 - \langle A \rangle^2 |\vec{b}|^2 -
\langle B \rangle^2 |\vec{a}|^2  & \geq &
(\vec{a} \cdot \vec{b})^2 - 2( \vec{a} \cdot \vec{b})
\langle A \rangle \langle B\rangle + \langle C \rangle^2\;. \nonumber
\end{eqnarray}
After rearranging the terms, we have
\begin{eqnarray}
|\langle A\rangle \vec{b} -
\langle B \rangle  \vec{a}|^2 + \langle C\rangle^2 \leq
|\vec{a}|^2| \vec{b}|^2 - (\vec{a} \cdot \vec{b})^2 = S^2
\;. \nonumber
\end{eqnarray}
The right hand side of the inequality is just the determinant of Gram
matrix of the vector $\vec{a}$, $\vec{b}$, which is the square of
area of parallelogram formed by $\vec{a}$, $\vec{b}$. The expectation
value is evaluated for certain quantum state which can be prepared by
projecting one partite of the bipartite entangled state onto specific
quantum state. For example, for the entangled state $|\psi_{12}
\rangle = \alpha |+\rangle_1|+\rangle_2 +
\beta|-\rangle_1|-\rangle_2$, by projecting the partite 2 onto a
specific state $|n_p^{+}\rangle_2 = \cos\frac{\theta}{2}|+\rangle +
e^{i\phi} \sin\frac{\theta}{2}|-\rangle$ (Eigenstate of
$\vec{\sigma}_2 \cdot \vec{n}_{p}$ where $\vec{n}_p =
(\sin\theta\cos\phi, \sin\theta\sin\phi, \cos\theta)$), we can get
arbitrary quantum state $|\psi_1^+\rangle$
\begin{eqnarray}
|\psi_1^+\rangle = \frac{ _2\langle n_p^+|\psi_{12}\rangle}{|\, _2\langle n_p^+ |
\psi_{12}\rangle|}  = \frac{1}{|\, _2\langle n_p^+|\psi_{12}\rangle|}
\left( \alpha \cos\frac{\theta}{2}| +  \rangle + e^{-i\phi} \beta
\sin \frac{ \theta}{2}|-\rangle \right)\; . \label{projector-bipartite}
\end{eqnarray}
Similar expression holds for $|\psi_1^-\rangle$ when projecting with
$|n_p^-\rangle_2$. The uncertainty relation holds for arbitrary state, so
for $|\psi_1^{\pm}\rangle$
\begin{eqnarray}
& &  |\langle A\rangle \vec{b} -
\langle B \rangle  \vec{a}|^2 + \langle C\rangle^2 \leq  S^2 \nonumber \\
& \Rightarrow & |\langle \psi_1^{\pm}| A_1 |\psi_1^{\pm} \rangle \vec{b} -
 \langle  \psi_1^{\pm}| B_1|\psi_1^{\pm} \rangle \vec{a} |^2 +
\langle  \psi_1^{\pm}| C_1| \psi_1^{\pm} \rangle^2 \leq S^2 \; . \label{Sch-pm}
\end{eqnarray}
Here the subscript $1$ standards for partite 1. Multiplying $|\,
_2\langle n_p^{\pm}|\psi_{12}\rangle|^2$ to Eq.(\ref{Sch-pm}) with
the corresponding superscript $\pm$ and adding the two inequalities
we have
\begin{eqnarray}
|_2\langle n_p^{+}|\psi_{12}\rangle|^2
|\langle \psi_1^+|A_1 |\psi_1^+\rangle \vec{b}  -
\langle  \psi_1^+| B_1 |\psi_1^+\rangle \vec{a} |^2 +
| _2\langle n_p^{+}| \psi_{12}\rangle|^2 \langle \psi_1^+| C_1 | \psi_1^+ \rangle^2  + & &  \nonumber \\
|_2\langle n_p^{-}|\psi_{12}\rangle|^2 |\langle \psi_1^-| A_1 | \psi_1^-\rangle \vec{b} -
\langle  \psi_1^-| B_1 |\psi_1^-\rangle \vec{a}|^2 +
| _2\langle n_p^{-}|\psi_{12}\rangle|^2 \langle \psi_1^-| C_1 | \psi_1^- \rangle^2
 & \leq & S^2 \; .
\end{eqnarray}
With Cauchy's inequality $\sum_{i}p_i\sum_{i}p_i a_i^2 \geq
(\sum_{i}p_ia_i)^2$, Eq.(\ref{projector-bipartite}), and the
following relation
\begin{eqnarray}
& & |_2\langle n_p^+|\psi_{12}\rangle|^2 |\langle \psi_1^+| A_1 |\psi_1^+\rangle| +
|_2\langle n_p^-|\psi_{12}\rangle|^2 |\langle \psi_1^-| A_1 |\psi_1^-\rangle| \nonumber \\
 & = & |\langle \psi_{12}|A_1 \otimes |n_p^+\rangle_2\langle n_p^+||\psi_{12} \rangle| +
 |\langle \psi_{12}|A_1 \otimes |n_p^-\rangle_2 \langle n_p^-||\psi_{12}\rangle| \nonumber \\
 & \geq & \left|\langle \psi_{12}| A_1 \otimes |n_p^+\rangle_2 \langle n_p^+||\psi_{12}\rangle -
  \langle \psi_{12}| A_1 \otimes |n_p^-\rangle_2 \langle n_p^-||\psi_{12}\rangle \right| \nonumber \\
  & = & \left|\langle \psi_{12}| A_1 \otimes (|n_p^+\rangle_2 \langle n_p^+| - |n_p^-\rangle_2\langle n_p^-|)
|\psi_{12}\rangle \right| \nonumber \\
  & = & \left|\langle \psi_{12}| A_1 \otimes P_2|\psi_{12}\rangle \right|
  = \left|E(A_1, P_2)\right| \; ,
\end{eqnarray}
we can get
\begin{eqnarray}
\left| E(A_1, P_2)\vec{b} - E(B_1, P_2)\vec{a} \right|^2 +
\left| E(C_1, P_2) \right|^2 \leq S^2 \; .
\end{eqnarray}
Q.E.D.

\section{Proof of Eq.(\ref{MDR-Correlation})} \label{appendix-2}

Proof of Eq.(\ref{MDR-Correlation}):
\begin{eqnarray}
& & |\vec{a}|^2 + |\vec{b}|^2 - (-1)^m [E(A_2,A_3) + E(B_1, B_2)] \nonumber \\
& = & \frac{1}{4} \left[ \epsilon^+(A)^2 + \eta^+(B)^2 +
\epsilon^-(A)^2 + \eta^-(B)^2 \right]
\; . \nonumber
\end{eqnarray}

\noindent {\bf Proof}: For the particular state $|\psi_1^{\pm}\rangle$,
taking the definitions of Eq.(\ref{project-to-psi1}), the measurement
precisions turn to
\begin{eqnarray}
|_2\langle n_p^{\pm}|\psi_{12}^{(m)} \rangle|^2\epsilon^{\pm}(A)^2 =
\langle \psi_3|\langle \psi_{12}^{(m)}| P_{2}^{\pm}
\left[ U_{13}^{\dag} (I_1 \otimes I_2 \otimes A_3) U_{13} -
A_1 \otimes I_2 \otimes I_3 \right]^2
P_{2}^{\pm}|\psi_{12}^{(m)}\rangle|\psi_3\rangle \; . \nonumber
\end{eqnarray}
The corresponding disturbances are
\begin{eqnarray}
|_2\langle n_p^{\pm}|\psi_{12}^{(m)} \rangle|^2 \eta^{\pm}(B)^2 =
\langle \psi_3|\langle \psi_{12}^{(m)}| P_{2}^{\pm}
\left[ U_{13}^{\dag} (B_1 \otimes I_2 \otimes I_3) U_{13} - B_1
\otimes I_2 \otimes I_3 \right]^2
P_{2}^{\pm}|\psi_{12}^{(m)}\rangle|\psi_3\rangle \; . \nonumber
\end{eqnarray}
Using the complete relation of projection operators, the summation of
the precision and disturbance for $|\psi_1^+\rangle$ and
$|\psi_1^-\rangle$ gives
\begin{eqnarray}
& & |\alpha_m|^2 \epsilon^+(A)^2 +
|\beta_m|^2 \epsilon^-(A)^2  \nonumber \\
&  = & \langle \psi_{3}|\langle \psi_{12}^{(m)}| \left[ U_{13}^{\dag}
(I_1\otimes I_2\otimes A_3) U_{13} -
A_1\otimes I_2 \otimes I_3 \right]^2 |\psi_{12}^{(m)}\rangle |\psi_{3}\rangle \; , \\
& & |\alpha_m|^2 \eta^+(B)^2 +
|\beta_m|^2 \eta^-(B)^2 \nonumber \\
& = & \langle \psi_{3}| \langle \psi_{12}^{(m)}| \left[ U_{13}^{\dag}
( B_1 \otimes I_2\otimes I_3) U_{13} -
B_1\otimes I_2 \otimes I_3 \right]^2 |\psi_{12}^{(m)}\rangle |\psi_{3}\rangle \; .
\end{eqnarray}
where $\alpha_m\equiv \ _2\langle n_p^+|\psi_{12}^{(m)}\rangle$, $\beta_{m}
\equiv \ _2\langle n_p^-|\psi_{12}^{(m)}\rangle$ and $|\alpha_m|^2 +
|\beta_m|^2=1$. Due to the properties of Eq.(\ref{rotation-invariant-12}),
we have
\begin{eqnarray}
& &  |\alpha_m|^2 \epsilon^+(A)^2 + |\beta_m|^2 \epsilon^-(A)^2 \nonumber \\
&  = & \langle \psi_{3}|\langle \psi_{12}^{(m)}| \left[ U_{13}^{\dag}
(I_1\otimes I_2\otimes A_3) U_{13} - (-1)^m
I_1 \otimes A_2 \otimes I_3 \right]^2 |\psi_{12}^{(m)}\rangle |\psi_{3}\rangle \; , \\
& & |\alpha_m|^2 \eta^+(B)^2 +
|\beta_m|^2 \eta^-(B)^2 \nonumber \\
& = & \langle \psi_{3}| \langle \psi_{12}^{(m)}| \left[ U_{13}^{\dag}
( B_1 \otimes I_2\otimes I_3) U_{13}
-(-1)^m I_1\otimes B_2 \otimes I_3 \right]^2 |\psi_{12}^{(m)} \rangle |\psi_{3}\rangle \; .
\end{eqnarray}
The measurement interaction only involves particles of $1$,$3$, thus it
commutates with operators acting on partite 2, so we have
\begin{eqnarray}
& & |\alpha_m|^2 \epsilon^+(A)^2 + |\beta_m|^2 \epsilon^-(A)^2 \nonumber \\
&  = & \langle \psi_{3}|\langle \psi_{12}^{(m)}| U_{13}^{\dag}
 \left(I_1\otimes I_2\otimes A_3 - (-1)^m
I_1 \otimes A_2 \otimes I_3 \right)^2
U_{13}|\psi_{12}^{(m)}\rangle |\psi_{3}\rangle \; , \label{pre-permute} \\
& & |\alpha_m|^2 \eta^+(B)^2 + |\beta_m|^2 \eta^-(B)^2 \nonumber \\
& = & \langle \psi_{3}| \langle \psi_{12}^{(m)}| U_{13}^{\dag}
\left( B_1 \otimes I_2\otimes I_3
-(-1)^m I_1\otimes B_2 \otimes I_3 \right)^2
U_{13} |\psi_{12}^{(m)} \rangle |\psi_{3}\rangle \; . \label{dis-permute}
\end{eqnarray}
Define $|\psi_{123}\rangle \equiv U_{13}|\psi_{12}^{(m)} \rangle
|\psi_3\rangle$, Eqs.(\ref{pre-permute},\ref{dis-permute}) turn to
\begin{eqnarray}
|\alpha_m|^2\epsilon^+(A)^2 + |\beta_{m}|^2\epsilon^-(A)^2
& = & \langle \psi_{123}| \left(A_3 - (-1)^{m} A_2 \right)^2
 |\psi_{123}\rangle \; , \\
|\alpha_m|^2 \eta^+(B)^2 + |\beta_m|^2 \eta^-(B)^2
& = & \langle \psi_{123}| \left( B_1 - (-1)^{m} B_2\right)^2  |\psi_{123}\rangle\; .
\end{eqnarray}
From the definition of operators $A=\vec{\sigma}\cdot \vec{a}$, $B =
\vec{\sigma}\cdot \vec{b}$, and the wave function $|\psi_{12}^{(m)}\rangle$
we have chosen (this gives $|\alpha_m|^2 = |\beta_m|^2 = 1/2$), the above
equations reduce to
\begin{eqnarray}
\frac{1}{2} \left[\epsilon^+(A)^2 + \epsilon^-(A)^2\right] =
2|\vec{a}|^2 - (-1)^{m}2E(A_2,A_3) \; , \\
  \frac{1}{2} \left[\eta^+(B)^2 + \eta^-(B)^2 \right] =
2|\vec{b}|^2 -(-1)^{m} 2 E(B_1,B_2) \; . \label{precision-disturbance}
\end{eqnarray}
This gives the relation Eq.(15). Q.E.D.

\section{Proof of Theorem \ref{Theorem-Heisenberg-Ozawa}} \label{appendix-3}

\noindent {\bf Proof}: Here we present the proof for $m=0$, the case of
$m=1$ can be derived similarly. For the Heisenberg-type MDR, taking $[A,B]
=2iC$ we have
\begin{eqnarray}
\epsilon^{+}(A)\eta^+(B) \geq |\langle \psi_1^+|C |\psi_1^+\rangle| \; , \;
\epsilon^{-}(A)\eta^-(B) \geq  |\langle \psi_1^-|C |\psi_1^-\rangle|\; . \nonumber
\end{eqnarray}
These hyperbolic form constraints on $\epsilon(A)$ and $\eta(B)$ with given
asymptotes are totally characterized by the distances from the vertices to
the origin of the coordinates. That is, the essence of the above
inequalities is characterized by
\begin{eqnarray}
\epsilon^{+}(A)^2 + \eta^+(B)^2 \geq 2 |\langle \psi_1^+|C|\psi_1^+\rangle| \; ,
\; \epsilon^{-}(A)^2 + \eta^-(B)^2 \geq 2 |\langle \psi_1^-|C|\psi_1^-\rangle| \; . \nonumber
\end{eqnarray}
The summation over the above two equations gives
\begin{eqnarray}
\epsilon^{+}(A)^2 + \eta^+(B)^2 + \epsilon^{-}(A)^2 + \eta^-(B)^2 \geq
2 (|\langle \psi_1^+|C|\psi_1^+\rangle| + |\langle \psi_1^-|C|\psi_1^-\rangle|)
\end{eqnarray}
The left hand side of the above inequality can be represented as correlation
functions via Eq.(15). The right hand sides of the inequality can be written
as
\begin{eqnarray}
&  & (|\langle \psi_1^+| C_1 |\psi_{1}^+\rangle| +
|\langle \psi_1^-| C_1 |\psi_{1}^-\rangle| )  \nonumber \\
& = & 2 \left( \left|\langle \psi_{12}^{(0)}| C_1 \otimes P_2^+|\psi_{12}^{(0)}\rangle \right| +
\left| \langle \psi_{12}^{(0)}| C_1 \otimes P_2^-|\psi_{12}^{(0)}\rangle \right|\right)
\nonumber \\ & \geq & 2 \left| \langle \psi_{12}^{(0)}|C_1 \otimes P_2^+|\psi_{12}^{(0)}\rangle -
\langle \psi_{12}^{(0)}|C_{1} \otimes P_2^-|\psi_{12}^{(0)}\rangle \right| \nonumber \\
& = & 2 \left|  \langle \psi_{12}^{(0)}|C_1 \otimes P_2|\psi_{12}^{(0)}\rangle \right|
\equiv 2 |E_{12}(C_1, P_2)| \; ,
\end{eqnarray}
where we have used Eq.(\ref{project-to-psi1}) and $P_2^{\pm} =
|n_p^{\pm}\rangle_2\langle n_p^{\pm}|$. It is clear that the essence of the
Heisenberg-type MDR, combining Eq.(\ref{MDR-Correlation}) and
Eq.(\ref{Heisenberg-MDR-PD}), is characterized by following inequalities
\begin{eqnarray}
E(A_2,A_3) + E(B_1,B_2) + |E_{12}(C_1,P_2)|
\leq |\vec{a}|^2 + |\vec{b}|^2\; . \label{choose-P2}
\end{eqnarray}
Here the bipartite correlation function $E_{12}$ is written with
subscript explicitly. Eq.(\ref{choose-P2}) must be satisfied for any
given $P_2$
\begin{eqnarray}
E(A_2,A_3) + E(B_1,B_2)
\leq |\vec{a}|^2 + |\vec{b}|^2 - |\vec{n}_p \cdot \vec{c}| \; .
\end{eqnarray}
This is just the Heisenberg upper bound for the correlations and its lower
limit is 0 for $m=0$.

From the Ozawa's MDR, we have
\begin{eqnarray}
& & \epsilon^{\pm}(A) \eta^+(B)+ \epsilon^{\pm}(A)\Delta^{\pm}(B) +
\eta^{\pm}(B) \Delta^{\pm}(A)\geq
|\langle \psi_1^{\pm} |C_1|\psi_{1}^{\pm}\rangle|  \nonumber \\
& \Rightarrow & \left[\epsilon^{\pm}(A) + \Delta^{\pm}(B) \right]
\left[\eta^+(B)+ \Delta^{\pm}(A)\right] \geq
|\langle \psi_1^{\pm} |C_1|\psi_{1}^{\pm}\rangle| + \Delta^{\pm}(A)\Delta^{\pm}(B)\; , \nonumber
\end{eqnarray}
where $\Delta^{\pm}(A,B)$ are the standard deviations evaluated with
$|\psi_1^{\pm}\rangle$. We see that the Ozawa's MDR is just a displaced
hyperbolic curve compared to the Heisenberg-type MDR. The characterization
distance of its vertices to the origin can be formulated as
\begin{eqnarray}
\epsilon^{\pm}(A)^2 + \eta^{\pm}(B)^2 \geq f[\Delta^{\pm}(A),\Delta^{\pm}(B),
|\langle \psi_1^{\pm} |C_1|\psi_{1}^{\pm}\rangle|] \; .
\end{eqnarray}
where $f$ is a function of $\Delta(A)$, $\Delta(B)$ and $|\langle
C\rangle|$. In order to make this inequality universally valid the left hand
side has to be greater than or equal to the maximum value of the right hand
side. Function $f$ gets the maximum value of $(2-\sqrt{2})^2|\langle
\psi_1^{\pm} |C_1|\psi_{1}^{\pm}\rangle|$ at $\Delta^{\pm}(A)^2 =
\Delta^{\pm}(B)^2 = |\langle \psi_1^{\pm} |C_1|\psi_{1}^{\pm}\rangle| $.
Similar as the case of Heisenberg-type MDR, we will get
\begin{eqnarray}
& & (|\vec{a}|^2 + |\vec{b}|^2) - \left[ E(A_2,A_3) + E(B_1, B_2) \right] \nonumber \\
& = & \frac{1}{4}( \epsilon^{+}(A)^2 + \eta^{+}(B)^2 +  \epsilon^{-}(A)^2 + \eta^{-}(B)^2)
\nonumber \\
& \geq & \frac{1}{2}(\sqrt{2}-1)^2 (|\langle \psi_1^+|C_1|\psi_1^+\rangle| +
|\langle \psi_1^-|C_1|\psi_1^-\rangle|) \nonumber \\
& \geq & (\sqrt{2}-1)^2 |E_{12}(C_1, P_2)|\; .
\end{eqnarray}
Thus the essence of the Ozawa's MDR is characterized by the following
inequalities
\begin{eqnarray}
E(A_2, A_3) + E(B_1, B_2) \leq |\vec{a}|^2 + |\vec{b}|^2 -
(\sqrt{2}-1)^2|\vec{n}_p \cdot \vec{c}| \; . \label{upper-Ozawa}
\end{eqnarray}
It should be noted here that the above constraint on correlations has no
lower limit because the MDRs (both Heisenberg-type and Ozawa's) does not
specify the upper limits. In the qubit systems, the upper bound for the
measurement precision and disturbance of the observables may be obtained
from the finite spectrums of the observable operators. Q.E.D.

\end{document}